\documentstyle[preprint,aps,epsf]{revtex}
\newif\iftightenlines\tightenlinesfalse
\tightenlines\tightenlinestrue

\def\eslt{\not\!\!{E_T}}
\def\to{\rightarrow}

\def\tu{\tilde u}

\def\tb{\tilde b}

\def\td{\tilde d}

\def\tst{\tilde t}
\def\ttau{\tilde \tau}

\def\tg{\tilde g}
\def\tnu{\tilde\nu}
\def\tell{\tilde\ell}

\def\tw{\widetilde W}
\def\tz{\widetilde Z}
\begin{document}
\draft
\preprint{\vbox{\baselineskip=14pt%
   \rightline{FSU-HEP-000714}
   \rightline{UH-511-969-00}
}}
\title{VIABLE SUPERSYMMETRIC MODELS WITH \\ 
AN INVERTED SCALAR MASS HIERARCHY \\
AT THE GUT SCALE}
\author{Howard Baer$^1$, Csaba Bal\'azs$^2$, Pedro Mercadante$^1$, 
Xerxes Tata$^2$ and Yili Wang$^2$}
\address{
$^1$Department of Physics,
Florida State University,
Tallahassee, FL 32306 USA
}
\address{
$^2$Department of Physics and Astronomy,
University of Hawaii,
Honolulu, HI 96822, USA
}
\date{\today}
\maketitle
\begin{abstract}

  Supersymmetric models with an inverted mass hierarchy (IMH:
  multi-TeV first and second generation matter scalars, and sub-TeV
  third generation and Higgs scalars) have been proposed to ameliorate
  phenomenological problems arising from flavor changing neutral
  currents (FCNCs) and CP violating processes, while satisfying
  conditions of naturalness. Models with an IMH already in place at
  the GUT scale have been shown to be constrained in that for many
  model parameter choices, the top squark squared mass is driven to
  negative values. We delineate regions of parameter space
  where viable models with a GUT scale IMH can be generated. 
  We find that larger values of GUT scale first and second generation
  scalar masses act to suppress third generation scalars, 
  leading to acceptable solutions if
  GUT scale gaugino masses are large enough. 
  We show examples of viable models
  and comment on their characteristic features. For
  example, in these models the gluino mass is bounded from below, and
  effectively decouples, whilst third generation scalars remain at
  sub-TeV levels. While possibly fulfilling criteria of naturalness,
  these models present challenges for detection at future $pp$ and
  $e^+e^-$ collider experiments.

\end{abstract}

\medskip



Supersymmetry offers an elegant solution to the problem of
quadratically  divergent scalar masses in the Standard Model (SM),
provided supersymmetric matter exists at or near the weak scale\cite{susy}.
A Minimal Supersymmetric Standard Model (MSSM) can be constructed, with
124 free parameters, most of which occur in the soft SUSY breaking (SSB) 
sector of the model\cite{dimop} and reflect our ignorance about how 
SUSY is broken. 
Taking arbitrary weak scale SSB parameter choices generally leads to
conflict with various low energy constraints associated with
flavor changing neutral currents (FCNCs), and CP violating processes 
such as the electric dipole moments of the proton and neutron\cite{masiero}. 
Of course, SUSY model builders have to explain the origin
of SSB terms while at the same time satisfying  constraints imposed by 
low energy processes.

Three possibilities have emerged for building models consistent with
low energy constraints: 
1. {\it universality} (degeneracy) of scalar masses\cite{georgi},
2. {\it alignment} of fermion and sfermion mass matrices\cite{ns} and 
3. {\it decoupling}, which basically involves setting sparticle masses
to such high values that SUSY loop effects are 
suppressed relative to SM loops\cite{dine}.
Models with gauge mediation\cite{nelson}, 
anomaly mediation\cite{randall} or gaugino 
mediation\cite{ss}
of SUSY breaking naturally lead to universality of 
particles with the same gauge quantum numbers. 
Supersymmetric models with SUSY breaking communicated via gravity
(supergravity models) in general lead to {\it non}-universal 
scalar masses\cite{kap}.
The minimal supergravity model (mSUGRA) adopts universality as an {\it ad hoc} 
assumption\cite{susy}. 
In this paper, we explore a class of
models which potentially solve 
the SUSY flavor and CP problems via the decoupling solution.

It is important to notice that 
``naturalness'' arguments\cite{nat}, which generally require sub-TeV
sparticle masses,  most directly apply
to third generation superpartners, owing to their large 
Yukawa couplings. In contrast, the constraints from flavor physics
mentioned above
apply (mainly) to scalar masses of just the first two generations.
This observation has motivated the construction of a variety of models,
collectively known as inverted mass hierarchy (IMH) models\cite{imh}, 
where the first and second generation squarks and sleptons have multi-TeV
masses, while third generation scalars have sub-TeV masses.
For models in which the IMH occurs at or near the GUT scale (GSIMH models),
it has been emphasized\cite{mur} that two loop contributions to 
renormalization group (RG) running can cause tachyonic third generation 
squark masses to occur, unless these masses are beyond $\sim 1$ TeV,
which again pushes the model towards the ``unnatural''.

Recently, it has been pointed out that models with a weak scale IMH can be 
generated radiatively by starting with multi-TeV scalar masses for all scalars
at $M_{GUT}$\cite{feng}.
For certain choices of GUT scale SSB boundary conditions, and assuming
Yukawa coupling unification, 
the Higgs and third generation SSB masses then evolve rapidly towards zero,
whilst first and second generation scalars remain heavy.
However, requiring realistic third generation fermion masses and also
a consistent radiative breakdown in electroweak symmetry (REWSB), only a rather
small IMH can be generated\cite{bmt}, 
which is  not sufficient by itself
to solve the SUSY flavor and CP problems.

An alternative approach to a decoupling solution is to take very large
values of scalar masses in models with intermediate to large values of
$\tan\beta$. The ``focus point'' behavior of the Higgs SSB masses 
results in models with {\it all} matter scalar above a TeV, 
but with low values of 
$|\mu |$, and possibly low fine-tuning\cite{fmm}. 
However, even in these models,
scalar masses are typically in the 1-3 TeV range, and are again not sufficient
to solve the SUSY flavor and CP problems without some degeneracy or alignment.

In this paper, we examine models with a scalar IMH already in place
at the GUT scale. 
We assume that the MSSM is a valid theory between $M_{GUT}$ and $M_{weak}$,
and that REWSB occurs.
As noted in Ref. \cite{mur}, it is crucial to work with two loop RGEs
for this class of models. The two loop RGEs for the MSSM have been
presented in \cite{mv}, and have been implemented in ISAJET versions
$\ge 7.49$\cite{isajet}.

We adopt the following parameter space for our studies:
\begin{equation}
m_0(1),\ m_0(3),\ m_{1/2},\ A_0,\ \tan\beta,\ sign(\mu ),
\end{equation}
where $m_0(1)$ is the common mass of all first generation matter scalars
at $M_{GUT}$, and $m_0(3)$ is the common third generation scalar mass.
For simplicity, we adopt $m_0(2)=m_0(1)$, although the whole point is that
$m_0(2)$ need not equal $m_0(1)$, 
so long as both are far above the TeV scale. The
Higgs scalar SSB masses are set equal to $m_0(3)$, and take values of
$\sim 1$ TeV. As usual, $m_{1/2}$ and $A_0$
are common GUT scale gaugino and trilinear masses, 
$\tan\beta ={{v_u}\over {v_d}}$, and $\mu$ is the superpotential Higgs
mass term.

For our renormalization group solution to the sparticle and Higgs mass 
spectrum, we use ISASUGRA (a part of the ISAJET package). Briefly, 
starting from weak scale values for the gauge and Yukawa couplings, ISASUGRA
evolves the couplings up in energy until the GUT scale is determined, where 
$g_1=g_2$. At $M_{GUT}$, the various SSB mass parameters are entered, and
the set of 26 coupled RGEs for gauge and Yukawa couplings, and SSB masses,
are evolved down to scale $M_{weak}$, where the sparticle mass spectrum
can be calculated. An iterative procedure is adopted so that the
RG improved one loop effective potential can be calculated and 
minimized at an optimized
scale choice $Q=\sqrt{m_{\tst_L}m_{\tst_R}}$, where REWSB is required.

The form of the two loop RGEs for SSB masses is given by\cite{mv}
\begin{eqnarray}
\frac{dm_i^2}{dt}=\frac{1}{16\pi^2}\beta_{m_i^2}^{(1)}+
\frac{1}{(16\pi^2)^2}\beta_{m_i^2}^{(2)},
\end{eqnarray}
where $t$ is the natural log of the scale,
$i=Q_j,\ U_j,\ D_j,\ L_j$ and $E_j$, and $j=1-3$ is a generation index.
Two loop terms are suppressed relative to one loop terms by
the square of a coupling constant, plus an additional factor of $16\pi^2$
in the denominator.
The two loop terms
\begin{equation}
\beta_{m_i^2}^{(2)}\ni a_ig_3^2\sigma_3+b_ig_2^2\sigma_2+c_ig_1^2\sigma_1,
\end{equation}
where 
\begin{eqnarray}
\sigma_1 &=& {1\over 5}g_1^2\{3(m_{H_u}^2+m_{H_d}^2)+Tr[{\bf m}_Q^2+
3{\bf m}_L^2+8{\bf m}_U^2+2{\bf m}_D^2+6{\bf m}_E^2]\},\\
\sigma_2 &=& g_2^2\{m_{H_u}^2+m_{H_d}^2+Tr[3{\bf m}_Q^2+
{\bf m}_L^2]\},\ \ \ \ {\rm and} \\
\sigma_3 &=& g_3^2Tr[2{\bf m}_Q^2+{\bf m}_U^2+{\bf m}_D^2],
\end{eqnarray}
and the ${\bf m}_i^2$ are squared mass matrices in generation space.
The numerical coefficients $a_i$, $b_i$ and $c_i$ are related to the quantum
numbers of the scalar fields, but are all positive quantities. Thus,
incorporation of multi-TeV masses for the first and second generation
scalars leads to an overall positive, {\it possibly dominant},
contribution to the slope of SSB mass trajectories versus energy scale.
Although formally a two loop effect, 
the smallness of the couplings is compensated by the 
much larger values of masses of the first two generations of scalars.
In running from $M_{GUT}$ to $M_{weak}$,
this results in an overall {\it reduction} of scalar masses, which is felt
most strongly by the sub-TeV third generation and Higgs scalar masses, and
indeed leads to the constraints found in Ref.~\cite{mur}. For values of
SSB masses which fall short of the constraints of Ref.~\cite{mur}, a sort of
see-saw effect amongst scalar masses occurs: the higher the value of
first and second generation scalar masses, the larger will be the 
two loop suppression of third generation and Higgs scalar masses, until
the constraint of Ref.~\cite{mur} takes effect (or the lightest SUSY particle
(LSP) ceases to be charge or color neutral).

An example of this effect is shown in Fig. \ref{imh3_1}, where we plot
in frame {\it a}) the evolution of third generation SSB scalar masses from 
a common GUT scale value of $m_0(3)=1000$ GeV, down to the weak scale.
We also take $m_0(1)=15$ TeV, $m_{1/2}=1400$ GeV, $A_0=0$, $\tan\beta =3$
and $\mu >0$.
The dashed curves represent the case for universal scalar masses,
with $m_0=1$ TeV for {\it all} scalars at $M_{GUT}$, while the solid lines
indicate the GSIMH model. 
We see that at scales close to $M_{GUT}$,
the trajectories of SSB scalar masses differ radically from the mSUGRA
case, due to the dominant two loop RGE contributions, and that furthermore,
these contributions overcome the positive one loop contributions from gauge 
interactions, and
actually suppress the scalar masses relative
to the mSUGRA case with universality. At lower energies, the one loop gauge 
contributions  to squark masses again become dominant--- 
due to increasing gluino mass and
$SU(3)$ gauge coupling--- resulting in an upward turn of the 
corresponding mass parameters.
However, the final weak scale values of SSB mass parameters 
in this case are suppressed by almost a factor of 2 relative to the model with
universality. In fact, most of the GSIMH model SSB masses are at sub-TeV
levels, as opposed to the multi-TeV scale of SSB masses from the mSUGRA
model. This means the GSIMH model could be in accord 
with naturalness constraints, even though the mSUGRA model is not.
A further interesting feature is that, since the two loop
RGE contributions to $m_L^2$ are larger than those to $m_E^2$, the lightest
third generation slepton is most likely to be dominantly {\it left}-handed,
instead of {\it right}-handed, as in the mSUGRA model.
In frame {\it b}), we see that the evolution of SSB Higgs masses 
is also strongly altered by two loop RGE contributions. The absolute
values of $m_{H_d}^2$ and $m_{H_u}^2$ are {\it diminished} relative
to the mSUGRA model case, again resulting in the GSIMH 
model being more natural.

The corresponding weak scale sparticle and Higgs mass spectra are shown in 
Table \ref{tab:cases} for the two cases shown in Fig. \ref{imh3_1}.
Two additional cases for a high value of $\tan\beta =35$ are also
shown. For the GSIMH model 
in case 1, the first and second generation scalar masses are all 
$\sim 15$ TeV, which should be large enough to suppress various flavor changing
and CP violating processes, with the possible exception of $K-\overline{K}$
system. Meanwhile, most of the third generation scalar masses are at
sub-TeV values, in accord with naturalness considerations.
The relatively high gluino mass ($m_{\tg}\sim 3.3$ TeV) means that
the gluino is effectively decoupling as well as first and second
generation scalars. It is a general feature of viable GSIMH models
that not only the gluino, but the other chargino and neutralino 
masses are constrained to be heavy.
\footnote{We recognize that the electroweak 
charginos and neutralinos have direct couplings to the Higgs bosons, 
and could result in a need for fine-tuning if they are too heavy.} 
This has important 
consequences for searches for GSIMH models at colliders. For instance,
a Next Linear Collider (NLC) $e^+e^-$ machine would need 
at least $1.5$ TeV in the 
CM frame to significantly access the lighter sparticles in this spectrum.
The last entry in the Table shows the overlap between $\ttau_1$ and $\ttau_L$
states, so that the lightest stau in the GSIMH model is predominantly 
left handed.
At the CERN LHC $pp$ collider, top and bottom squark pair
production will be the dominant SUSY process. 
We have generated collider events for all SUSY production processes
for case 1 GSIMH model using ISAJET. We examined the various multi-jet
plus multi-isolated lepton plus missing $E_T$ signals using the
standard cuts and SM backgrounds given in Ref. \cite{lhc}.
In none of the signal channels examined was case 1 GSIMH model
visible above SM background at a $5\sigma$ level, assuming 10 fb$^{-1}$
of integrated luminosity.
We also looked at possible signals in multijet $+\eslt$ events with 
two tagged $b$-jets.
We required the cuts of Ref. \cite{gmsblhc}: $\eslt > 100$ GeV,
$p_T(b-jets)> 50$ GeV, and $\eslt +\sum E_T(jets)> 1500$ GeV (where the 
sum runs over non-tagged jets). In addition, the highest $E_T$ $b$-jet
was required to have $p_T> 100$ GeV. Again, for this case, no signal was 
visible above SM background for 10 fb$^{-1}$.
Alternatively, examination of the ``effective mass'' distribution has been
advocated as a means to quickly establish the presence of a SUSY signal
against SM backgrounds\cite{frank}. The effective mass is defined as
\begin{equation}
M_{eff}=E_T(j1)+E_T(j2)+E_T(j3)+E_T(j4)+\eslt ,
\end{equation}
where $j1$ refers to the highest $E_T$ jet in the event, and so forth.
In Fig. \ref{meff1}, we plot the effective mass signal for case 1
(open circles) and backgrounds taken from Ref. \cite{frank},
after using the cuts of \cite{frank}. For the five mSUGRA case studies of
Ref. \cite{frank}, the signal always emerges from the background 
$M_{eff}$ distribution at a suitably high value of $M_{eff}$; for our
GSIMH case 1, this does not happen.
For the spectra of the GSIMH model case 1, clearly a more clever
dedicated set of cuts will be needed to establish a SUSY signal.
Finally, we note that the spectra of GSIMH case 1 may well be excluded by
constraints from the cosmological relic density of neutralinos\cite{relic}, 
since
dominant neutralino annihilation likely takes place via $t$-channel
stau, sbottom and stop exchange, and these particles are rather heavy.
However, since $m_{\tz_1}$ is getting close to $m_{\ttau_1}$ and $m_{\tst_1}$,
co-annihilation effects may be important, and could work to reduce
the relic density to acceptable levels\cite{falk}.

In case 2, with $\tan\beta =35$, we take $m_0(1)=m_0(2)=10$ TeV, 
$m_0(3)=900$ GeV, $m_{1/2}=1000$ GeV and $A_0=0$. Again, the third 
generation scalars are generally suppressed to sub-TeV values compared
to their mSUGRA counterparts. The top squarks and staus are the lightest
third generation scalars, with the $\ttau_1$ again
predominantly left handed. 
We also examined GSIMH model case 2 for visibility at the LHC just as 
for case 1. Again, we found no observable signal using the simple cuts of 
Ref. \cite{lhc} or for the special $b$-jets cuts listed above.
The $M_{eff}$ distribution is shown in Fig. \ref{meff2}. As with case 1,
signal is always below background levels.
The spectra of GSIMH case 2 may well be allowed by constraints
from the cosmological neutralino relic density, since
$m_{\tz_1}\sim {m_A\over 2}$, and the neutralinos can annihilate efficiently
through the very wide $s$-channel Higgs graphs to $b\bar{b}$ final states.

To map out the allowable parameter space range for viable GSIMH models, 
we performed a scan over parameter space, generating random samples
of input parameters over the following ranges:
\begin{eqnarray*}
m_0(1)=m_0(2)&:& 2000\to 25,000\ {\rm GeV},\\
m_0(3)       &:& 0\to 2000\ {\rm GeV},\\
m_{1/2}      &:& 0\to 2000\ {\rm GeV},\\
A_0          &:& -10,000\ {\rm GeV}\to +10,000\ {\rm GeV},\\
\tan\beta    &:& 3\to 55,\ \ {\rm and}\\
\mu          &:& +\ {\rm or}\ - .
\end{eqnarray*}
In Fig. \ref{imh3_2}, each dot represents a viable model. Models are
ruled out if {\it i.}) REWSB does not occur, {\it ii.}) 
masses for the {\it physical} scalar eigenstates become tachyonic, or
{\it iii.})  the $\tz_1$ is not the LSP. 
In all frames, the value of $m_0(1)$ is 
plotted on the vertical axis. In {\it a}), we plot models versus $m_0(3)$. 
It is seen that larger values of $m_0(1)$ require
larger values of $m_0(3)$.
In fact, this is just the result of Ref. \cite{mur}.
In frame {\it b}), we plot $m_0(1)$ versus $m_{1/2}$.
Our main point here is that
for large enough values of $m_0(1)$, there is a minimum value of $m_{1/2}$
that is required: large values of gaugino 
masses give large {\it negative} slope
contributions to SSB scalar mass trajectories, which can off-set the large
positive two loop contributions refered to earlier. In general, the lower
bound on $m_{1/2}$ leads to rather large gaugino masses at the weak scale, 
which makes detection of SUSY at colliders more difficult, even though
third generation scalars can remain at sub-TeV values.
In frame {\it c}), we plot $m_0(1)$ versus the parameter $A_0$, 
and see that wide 
ranges of the $A_0$ parameter give rise to viable GSIMH models. Finally, in
{\it d}), we plot $m_0(1)$ versus $\tan\beta$, and see that no range of
$\tan\beta$ is particularly favoured.

In Fig. \ref{imh3_3}, we show sparticle masses and $\mu$ contours in the
$m_0(3)\ vs.\ m_0(1)$ plane, where we take $m_{1/2}=500$ GeV in frames
{\it a}) and {\it b}), and $m_{1/2}=1000$ GeV in frames {\it c}) and {\it d}).
We take $A_0=0$ and $\mu >0$ in all frames, 
and $\tan\beta =3$ in {\it a}) and {\it c}),
and $\tan\beta =35$ in {\it b}) and {\it d}). In the dark shaded regions
$\mu^2 < 0$ so that REWSB fails, 
while the light shaded regions have a non-neutralino LSP;
the intermediate shaded region has tachyonic scalar masses.
In unshaded regions viable models may be possible. 
Some authors may then impose further constraints 
from fine-tuning, FCNC or other considerations. 
The constraints that we delineate are the minimal
constraints any such model should satisfy.
We also show mass contours 
of 1 TeV for $\ttau_1$, $\tb_1$, $\tst_1$ and the $\mu$ parameter.
The labels are located in the regions with mass less than 1 TeV. An 
exception occurs in frame {\it b}), where $\mu$ is always less than 1 TeV; 
there we plot a contour of $\mu =300$ GeV. In frames {\it a}) and {\it b}),
with the smaller values of $m_{1/2}$, the magnitude of $m_0(1)$ is
generally limited, especially if one requires sub-TeV third generation
scalar masses, which occur for lower values of $m_0(3)$.
Moving to higher $m_{1/2}$ values in {\it c}) and {\it d}), significantly
larger values of $m_0(1)$ are allowed, which translates to greater suppression
of FCNC and CP violating processes. Even so, sub-TeV third generation scalar
masses are still possible in significant parts of the parameter plane. 
In the large
$\tan\beta$ plots, we see that $m_0(3)$ is restricted from above by 
REWSB. 

In Fig. \ref{imh3_4}, we show potentially viable regions 
in the $m_0(3)\ vs.\ m_{1/2}$
plane, for $m_0(1)=m_0(2)=10$ TeV, $A_0=0$, $\mu >0$ 
and {\it a}) $\tan\beta =3$ and {\it b}) $\tan\beta =35$. 
Points with open circles are excluded by REWSB, those with stars
by tachyonic masses and finally those with dots have a non-neutralino LSP. 
In addition,
various sparticle mass contours are shown. All contours show sparticle masses
of 1 TeV, except the $\tg$ contour shows $m_{\tg}=3$ TeV, and the $\tz_1$
contour shows $m_{\tz_1}= 500$ GeV. We see that for a fixed value of 
$m_0(1)$ , low values of $m_0(3)$ give rise to sub-TeV stau masses, while
low values of $m_{1/2}$ give rise to sub-TeV top and bottom squark
masses. 
The lightest neutralino is gaugino-like throughout the frame of
Fig. \ref{imh3_4}{\it a}, which in general can lead to large values for
the relic density of neutralinos, except for regions
where co-annihilation effects are important (along the boundary
of the excluded region).

At this point, some comparison with the results of Arkani-Hamed and Murayama
(AM) and Agashe and Graesser (AG) of Ref. \cite{mur} seems worthwhile.
As an example, taking $m_0(1)=10$ TeV, and $m_{1/2}=500$ GeV,
then AM (Fig. 2) require $m_0(3)>1500$ GeV, AG (Fig. 4a) 
require $m_0(3)>1100$ GeV
and we require, from Fig. \ref{imh3_3} for $\tan\beta =3$,
$m_0(3)>2400$ GeV. Taking instead $m_{1/2}=1000$ GeV, AM and AG require
$m_0(3)>300$ GeV, while we require $m_0(3)>600$ GeV.
The work of AM neglects all Yukawa couplings in the RGE evolution, 
and requires only positive SSB scalar squared masses; the work of AG
is similar, but takes into account the top quark Yukawa coupling.
In our work, we include the top, bottom and tau Yukawa couplings in 
the two loop RGEs. Our constraints are somewhat different, also,
since we require no tachyonic {\it physical} masses, a neutralino LSP, 
and REWSB, while the authors of Ref.\cite{mur} 
include some fine-tuning constraints. We remark that the exact location of
the boundary of the tachyonic excluded region can be very sensitive to
the manner in which the superparticle mass spectrum is calculated.
For instance, using ISAJET, the tachyonic excluded region occurs
during the first pass of RG evolution, when sparticle masses are calculated 
at tree level. The top quark Yukawa coupling 
can be significantly larger during the first iteration than the second, since
loop corrections are not yet included. This can cause a greater suppression
of top squark squared masses than in a fully consistent one loop
treatment, for which the bounds on $m_0(3)$ would be somewhat lower.
In a fully consistent one loop calculation of sparticle masses, the excluded
region should always come from the LSP and REWSB constraints, instead
of from tachyonic masses.

{\it Summary and Conclusions:}
We have investigated supersymmetric models with a scalar IMH in place
at the GUT scale. Two loop RGEs for 
SSB masses are crucial for this analysis. 
We map out parameter regions with first and
second generation scalar masses in the 5-20 TeV range but with sub-TeV
third generation scalars. These models go 
a long way towards solving the SUSY flavor and CP problems, while
remaining on the edge of naturalness. 
At the very least, all models of this type
will have to satisfy the minimal constraints that we have imposed.
Finally, since the GUT scale gaugino mass
is bounded from below, charginos, neutralinos and gluinos are generally rather
heavy, making the resulting sparticle mass spectra challenging to
discover at planned future collider facilities.

%
\acknowledgments
This research was supported in part by the U.~S. Department of Energy
under contract number DE-FG02-97ER41022 and DE-FG-03-94ER40833.
%

%

%
\newpage
%
%

\iftightenlines\else\newpage\fi
\iftightenlines\global\firstfigfalse\fi
\def\dofig#1#2{\epsfxsize=#1\centerline{\epsfbox{#2}}}

\begin{table}
\begin{center}
\caption{Weak scale sparticle masses and parameters (GeV) for two cases of
mSUGRA and GSIMH models.}
\bigskip
\begin{tabular}{lccccc}
\hline
parameter & mSUGRA & GSIMH & mSUGRA & GSIMH \\
          & case 1 & case 1 & case 2 & case 2 \\
\hline

$m_0(1)$ & 1000.0  & 15000.0 & 900.0 & 10000.0 \\
$m_0(3)$ & 1000.0  & 1000.0 & 900.0  & 900.0 \\
$m_{1/2}$ & 1400.0  & 1400.0 & 1000.0  & 1000.0 \\
$A_0$    & 0.0     & 0.0    & 0.0 & 0.0 \\
$\tan\beta$ & 3 & 3 & 35 & 35 \\
$m_{\tg}$ & 2965.1  & 3291.6 & 2181.7 & 2395.4 \\
$m_{\tu_L}$ & 2745.6 & 14993.0 & 2076.5 & 10014.8 \\
$m_{\td_R}$ & 2639.9 & 15009.0 & 1997.4 & 10024.2 \\
$m_{\tell_L}$ & 1351.8 & 14983.5 & 1110.7 & 9991.0 \\
$m_{\tell_R}$ & 1122.3 & 14991.4 & 970.9 & 9994.9 \\
$m_{\tnu_{e}}$ & 1349.9 & 14983.4 & 1107.8 & 9990.7 \\
$m_{\tst_1}$& 2131.2 & 658.7 & 1616.2 & 792.6 \\
$m_{\tst_2}$& 2546.4 & 1113.7 & 1877.8 & 1072.2 \\
$m_{\tb_1}$ & 2527.2 & 922.3 & 1834.7 & 940.0 \\
$m_{\tb_2}$ & 2638.8 & 1316.6 & 1907.6 & 1147.2 \\
$m_{\ttau_1}$ & 1121.0 & 671.6 & 835.4 & 690.0 \\
$m_{\ttau_2}$ & 1351.4 & 851.0 & 1063.7 & 761.2 \\
$m_{\tnu_{\tau}}$ & 1349.4 & 667.9 & 1055.3 & 721.4 \\
$m_{\tw_1}$ & 1110.7 & 1097.0 & 785.3 & 745.2 \\
$m_{\tz_2}$ & 1110.6 & 1097.7 & 785.2 & 745.8 \\
$m_{\tz_1}$ & 600.3 & 616.5 & 424.6 & 432.1 \\
$m_h$ & 112.8 & 111.8 & 122.4 & 126.5 \\
$m_A$ & 2302.3 & 1447.7 & 1218.3 & 801.9 \\
$m_{H^+}$ & 2305.2 & 1450.7 & 1222.5 & 807.7 \\
$\mu$ & 1721.8 & 1200.2 & 1070.0 & 792.3 \\
$\langle\tau_L|\tau_1\rangle$ & 0.02 & 0.99 & 0.15 & 0.71 \\
\hline
\label{tab:cases}
\end{tabular}
\end{center}
\end{table}

\newpage
%

%
\begin{figure}
\dofig{4in}{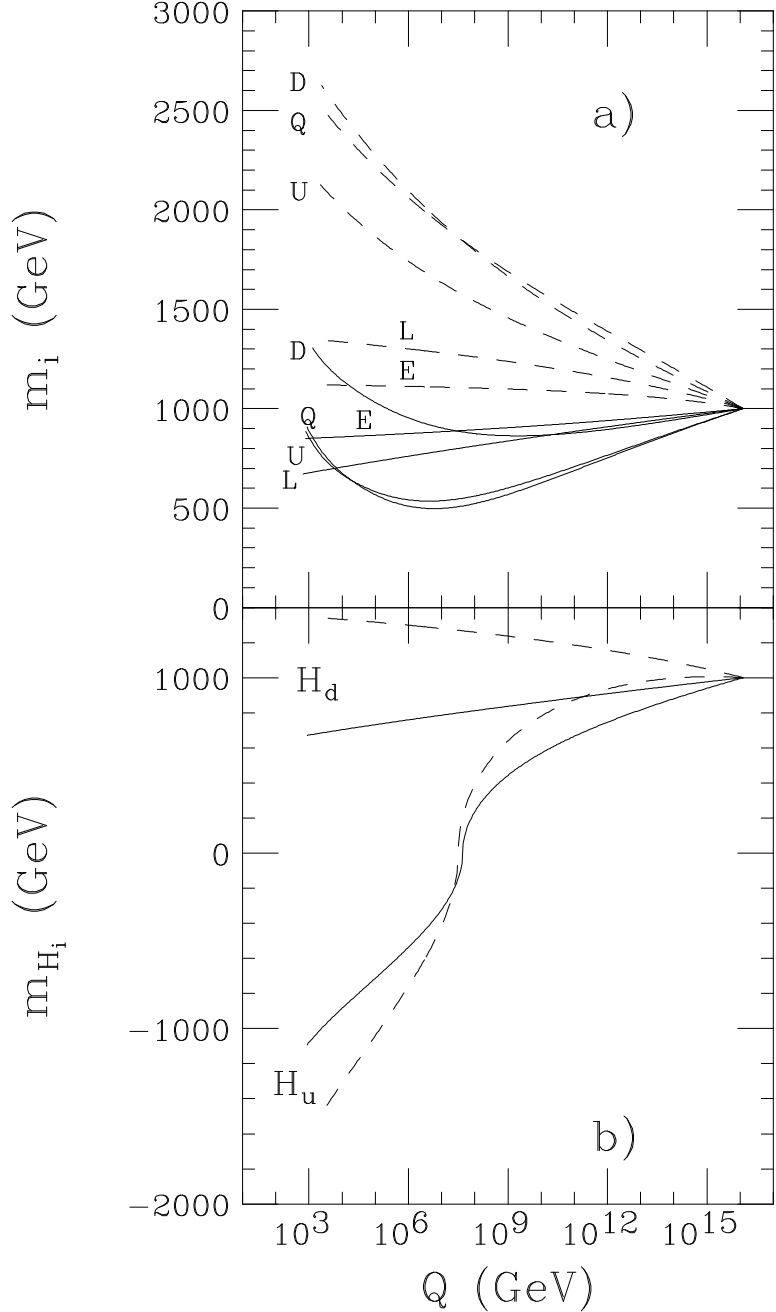}
\caption[]{
Evolution of {\it a}) third generation SSB masses 
and {\it b}) Higgs SSB masses from $M_{GUT}$ to $M_{weak}$ versus
scale choice $Q$. Dashed lines indicate mSUGRA model while solid lines
indicate the GSIMH model. We take $m_0(3)=1000$ GeV, 
$m_{1/2}=1400$ GeV, $A_0=0$, $\tan\beta =3$ and $\mu >0$.
In the GSIMH model, $m_0(1)=m_0(2)=15,000$ GeV, while in mSUGRA,
$m_0(1)=m_0(2)=m_0(3)$.}
\label{imh3_1}
\end{figure}
\begin{figure}
\dofig{5in}{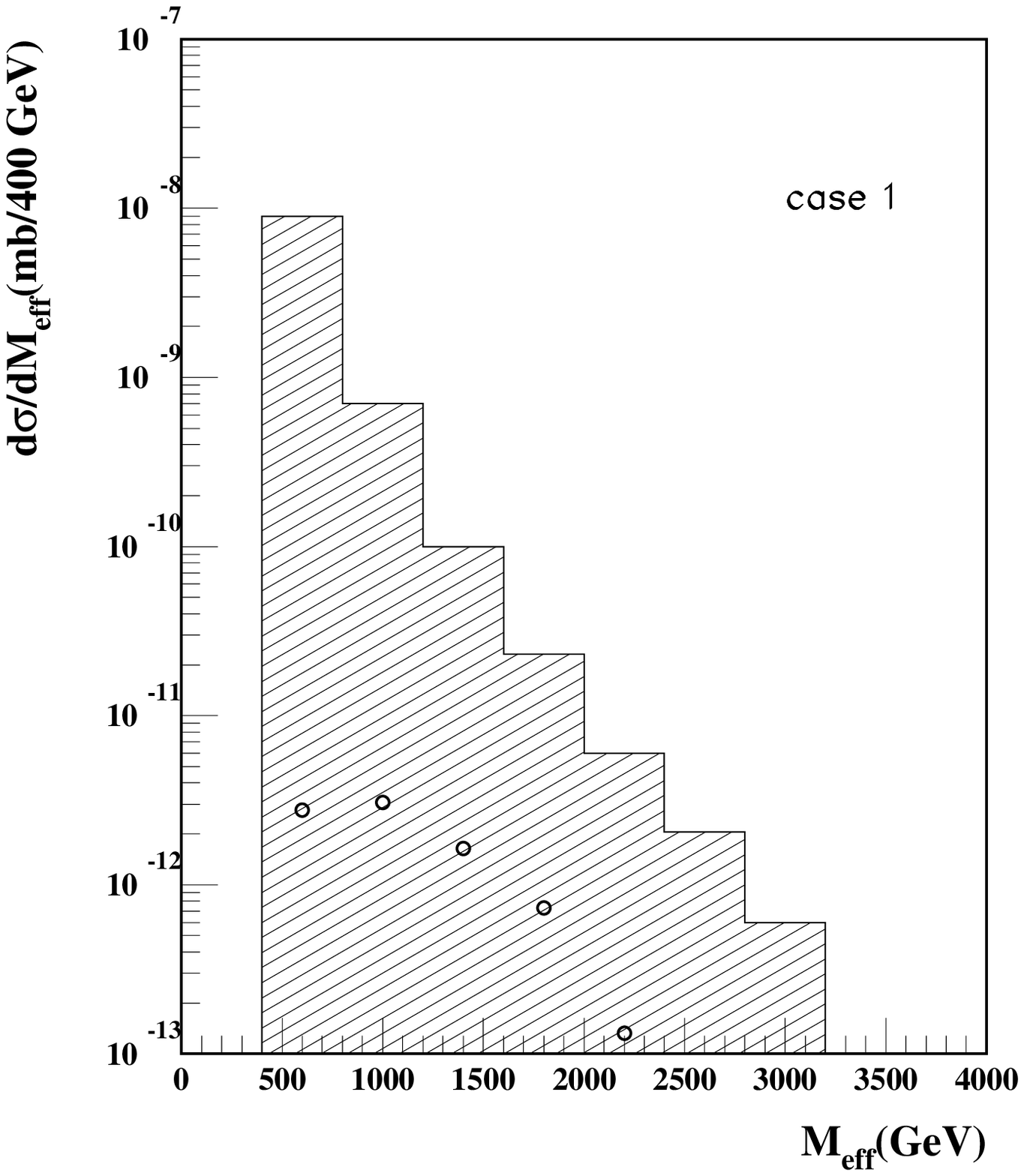}
\caption[]{A plot of the effective mass distribution background (histogram),
and signal (open circles) for GSIMH model case 1.}
\label{meff1}
\end{figure}
\begin{figure}
\dofig{5in}{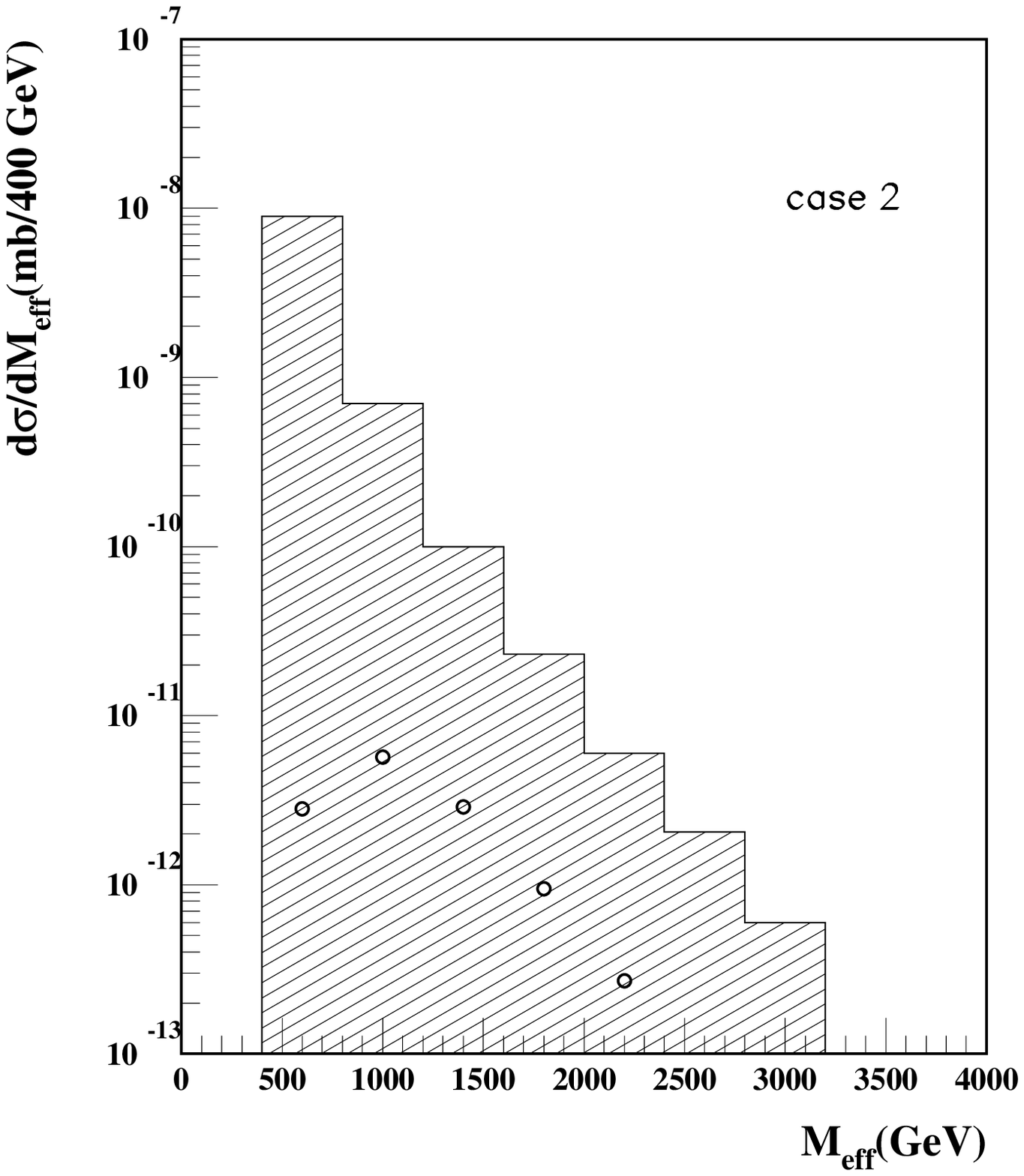}
\caption[]{A plot of the effective mass distribution background (histogram),
and signal (open circles) for GSIMH model case 2.}
\label{meff2}
\end{figure}
\begin{figure}
\dofig{5in}{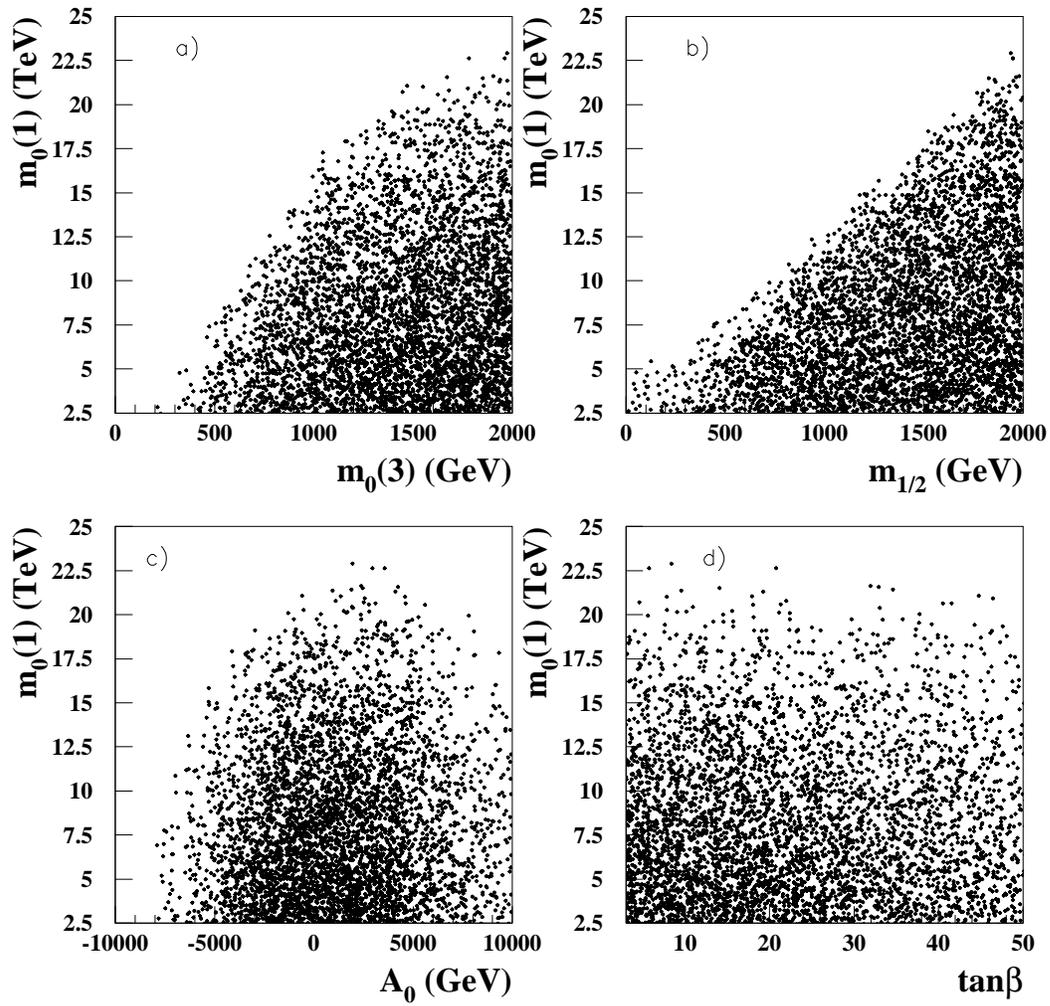}
\caption[]{A plot of parameter space where viable GSIMH models 
are located. The vertical axis is $m_0(1)$, while in 
{\it a}) models are plotted versus $m_0(3)$, in {\it b}) versus
$m_{1/2}$, in {\it c}) versus $A_0$ and in {\it d}) versus $\tan\beta$.
}
\label{imh3_2}
\end{figure}
\begin{figure}
\dofig{5in}{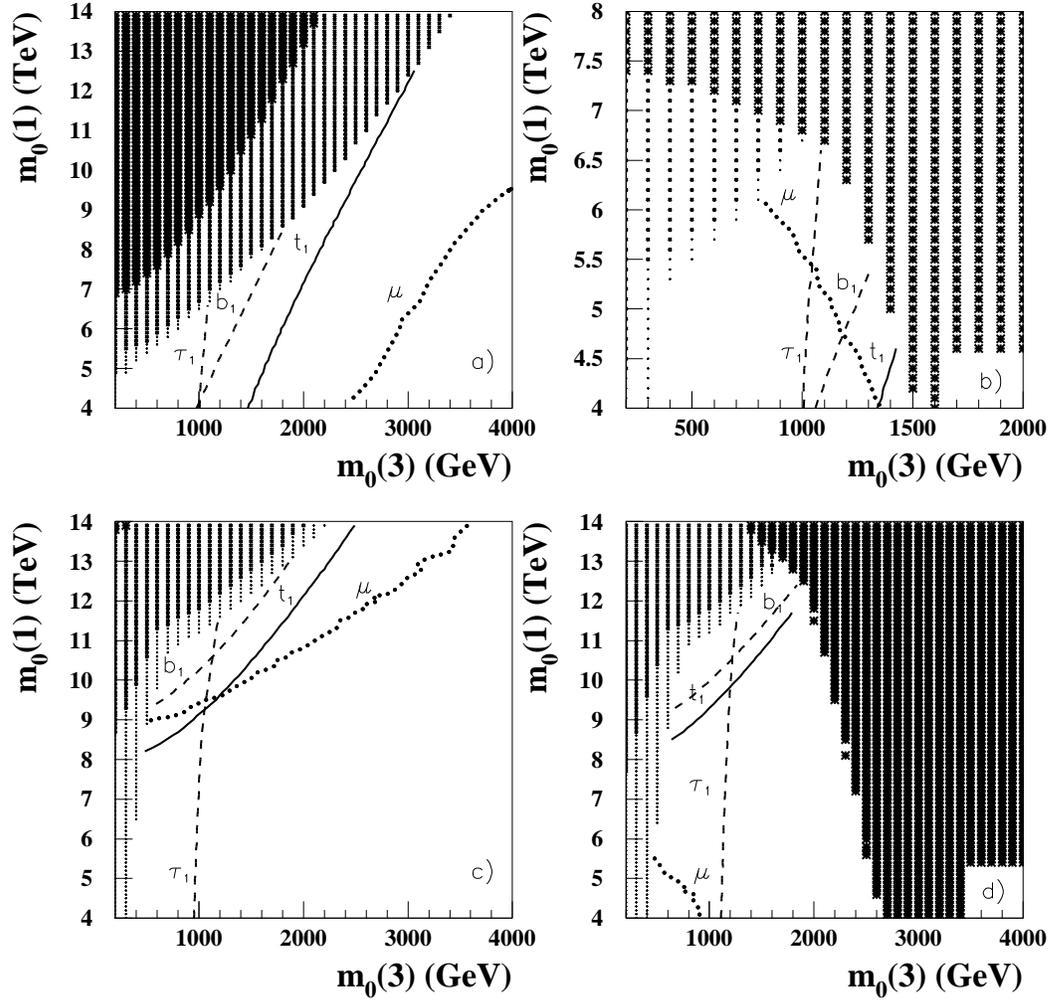}
\caption[]{A plot of the $m_0(3)\ vs.\ m_0(1)$ plane 
where viable GSIMH models are located. 
In {\it a}) and {\it b}), we take $m_{1/2}=500$ GeV, while in {\it c})
and {\it d}, we take $m_{1/2}=1000$ GeV. 
In {\it a}) and {\it c}), $\tan\beta =3$, while in {\it b})
and {\it d}, $\tan\beta =35$. We take $A_0=0$ and $\mu >0$. 
The contours show where various sparticle masses and the $\mu$ parameter
are equal to $1000$ GeV (except frame {\it b}), where $\mu =300$ GeV).
The darkest region is excluded by lack of REWSB, while the lightest
excluded region has a non-neutralino LSP. Tachyonic sparticle masses
are generated in the intermediate shaded regions.
}
\label{imh3_3}
\end{figure}
\begin{figure}
\dofig{5in}{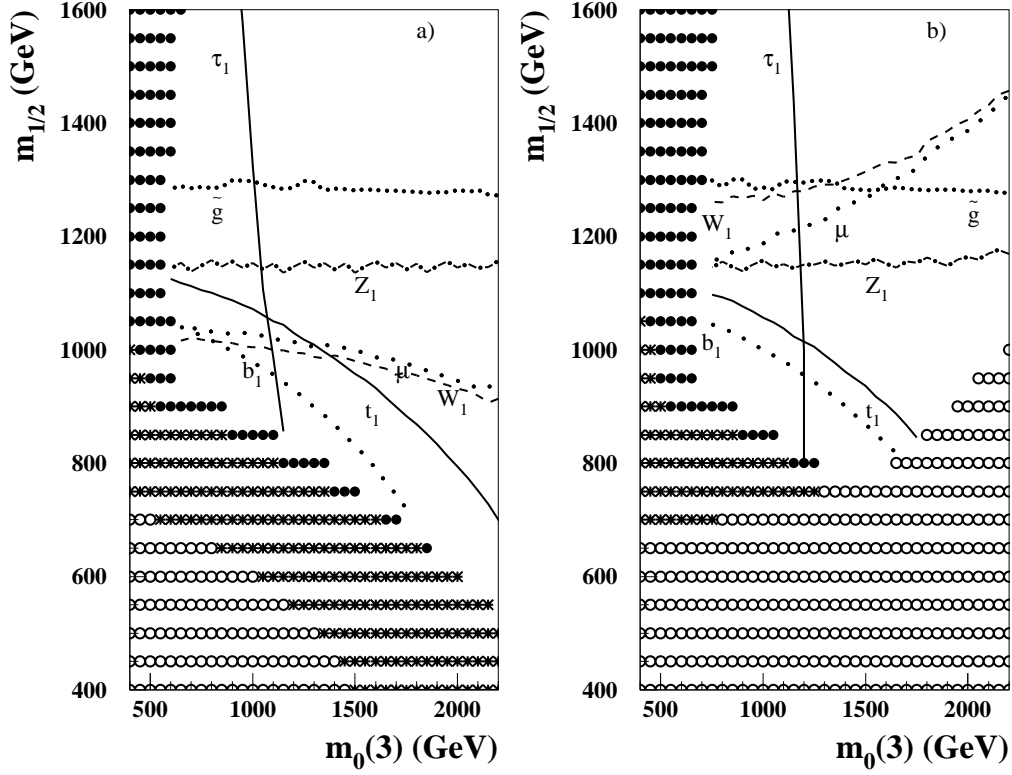}
\caption[]{A plot of parameter space where viable GSIMH models 
are located in the $m_0(3)\ vs.\ m_{1/2}$ plane, where we
take $m_0(1)=m_0(2)=10$ TeV, $A_0=0$, $\mu >0$ and 
{\it a}) $\tan\beta =3$ and {\it b}) $\tan\beta =35$.
Points with circles have no REWSB, points with stars have tachyonic masses and 
dots represent points with a charged and/or colored LSP. All contours represent
sparticle masses of 1 TeV, except $m_{\tg}<3$ TeV below the $\tg$ contour,
and $m_{\tz_1}<500$ GeV below the $\tz_1$ contour.
}
\label{imh3_4}
\end{figure}

\end{document}